# An Anomalous Contrast of Insulators in Scanning Electron Microscope: The Split Image

Lilin Xie[1], Xiaona Zhang[1]

[1]*Institute of Microstructure and Property of Advanced Materials, Beijing University of Technology, Beijing 100124, People's Republic of China*

**Abstract:** A novel phenomenon of anomalous contrast in scanning electron microscope when the instrument is used to observe an insulator specimen with a wolfram probe, we called double imaging, is reported in this article. We give a detail analysis of this phenomenon in its occurrence, and discuss the influence of the added probe to internal electric field which lead to the occurrence of double imaging.

**Introduction**

When uncoated insulating materials are observed in a scanning electron microscope (SEM), charging effect occurred and a distorted image will be formed[1]. The electric field in the vacuum resulting from the trapped electrons in the specimen leads to an image distortion owing to secondary electrons (SE) contrast's dependence on the local sample charging. One of the most spectacular observations related to the negative charging is the well-known mirror effect[2], and some other phenomenon caused by charging effect are also reported[3, 4]. The formation of the contrast is the result of a two-step process. The first step is the charging up of part of the surface of the specimen until it acts as an electron mirror with sufficient strength to deflect the electrons of the primary electron beam towards the electron collector. The second step is the building up of an image of the detector as the reflected beam scans over it, and then anomalous image appeared on screen.

The charging effect of electron bombardment on insulating materials is an important phenomenon which affects scanning electron microscope imaging[5-8]. When insulating sample is subjected to electron irradiation, primary incident electrons are more than outgoing ones (backscattered and secondary electrons), in the case of insulators this excess is trapped in the sample and resulting in the charging effect. The negative trapped charge in the sample lead to a surface potential, and the insulating material become negatively charged, giving rise to an electric field large enough to induce an electric breakdown in the insulator or to cause electrostatic deflection of the incident electron beam itself.

In this article, a novel phenomenon observed in SEM is reported and an explanation for the formation of a new type distort contrast is proposed. The experiment is carried out in SEM, different from any other anomalous contrast reported before, a same contrast of the sample which is called double imaging occurred when a metal probe approaching the insulating sample. The explanation is focus on the charging effect and the influence of metal probe in internal electric field. Charging effect lead to a surface potential on the sample, the metal probe changes the internal electric field and cause electrostatic deflection of the primary incident electron beam, thus another same sample contrast appeared, the experimental and theoretical results are also analyzed.

**Experimental Results**

Experiments were carried out in a SEM (FEI), the sample used in the experiment was $SiO_2$@Ag particle synthesized by thermal evaporation ton glass substrate. Fig. 1 is the typical SEM image of $SiO_2$@Ag particles with a common model, the structure of these particles have been studied before: Ag fill up the hemispherical structure and $SiO_2$ outcrop as pellet. Since $SiO_2$ and glass act as substrate are all insulating material, charging effect can be seen in the Figure 1(a) as there is an anomalous bright contrast.

Another experiment was carried out in SEM equipped with a cathode fluorescent (CL) spectrometer. Figure 1(b) is the image of the same particle in CL model, corresponded bright spots are $SiO_2$ outcrop, obviously, charging effect is occurred but no significant difference than before.

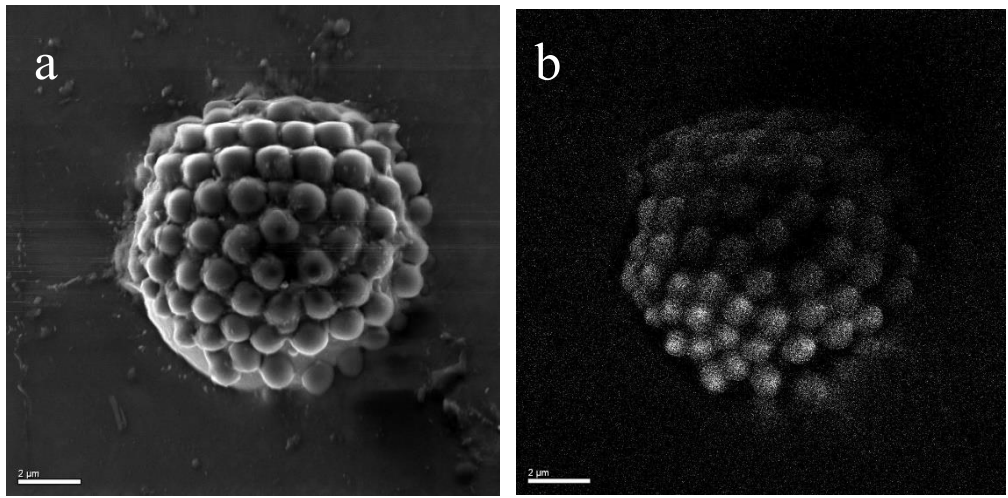

Fig. 1. (a)SEM image of SiO$_2$@Ag particle; (b)corresponding CL model image.

To have a further study of the charging effect of the particles in SEM, another experiment was carried out in SEM with a wolfram probe. The wolfram probe was ground connection and fix on a workbench which can do controlled planar and vertical movement. When the probe was controlled next to the particle by planar movement, the image become anomalous as shown in Figure 2. It is known that charging effect will lead to anomalous contrast occur in insulator materials, however, in this experiment, the anomalous contrast image of particle in SEM generate a double image when the probe across the sample, this phenomenon have never been report before. From Figure 2, it can be clearly seen that the image changed from distort to a double image with the probe approach. This phenomenon was not simply caused by charging effect, moreover, the double image of sample was a duplication instead of mirroring as shown in Figure 2 (d).

To understand this phenomenon, further experiment was carried out, the probe in the experiment before shown in Figure 2 was approaching in planar, in the next experiment the probe change in vertical height as shown in Figure 3. The phenomenon of these two experiment were significantly different from the phenomenon occurred in Figure 1, and the only change was the added metal probe which is the key to this phenomenon.

Series of recorded micrographs are presented in Figure 2. The contrast begin to slightly distort when the probe approaching the particles as shown in Figure 2 (a), as the probe is approaching closely the contrast become anomalous (b)(c) until split into double image (d).

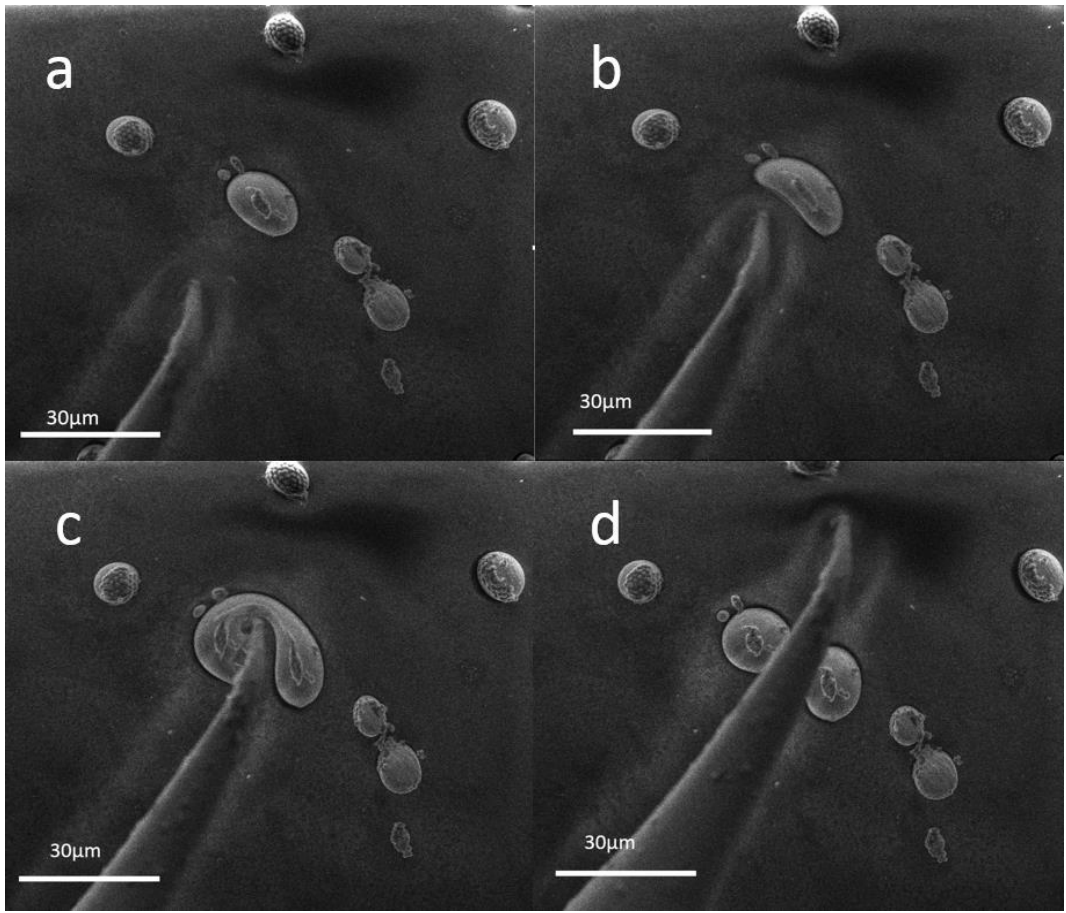

Fig .2 .SEM image with a wolfram probe.
From (a) to (d): the wolfram probe approach and across particle.

In the Fig. 3, charging effect is also occurred, micrographs recorded with the probe changed in vertical height. As shown in Figure 3 (a), there is no evident anomalous contrast in sample when the probe touched sample (the pinpoint of probe was bend as it hit the sample accidentally). However, with the height change, the image of the sample become anomalous and become a double image after probe away from sample in some height, with the height increase the contrast become unclearly at last. Figure 2 and Figure 3 show two process of double imaging, this phenomenon present in a charging condition and generated by the wolfram probe, obviously, probe is the key requirement to the double imaging.

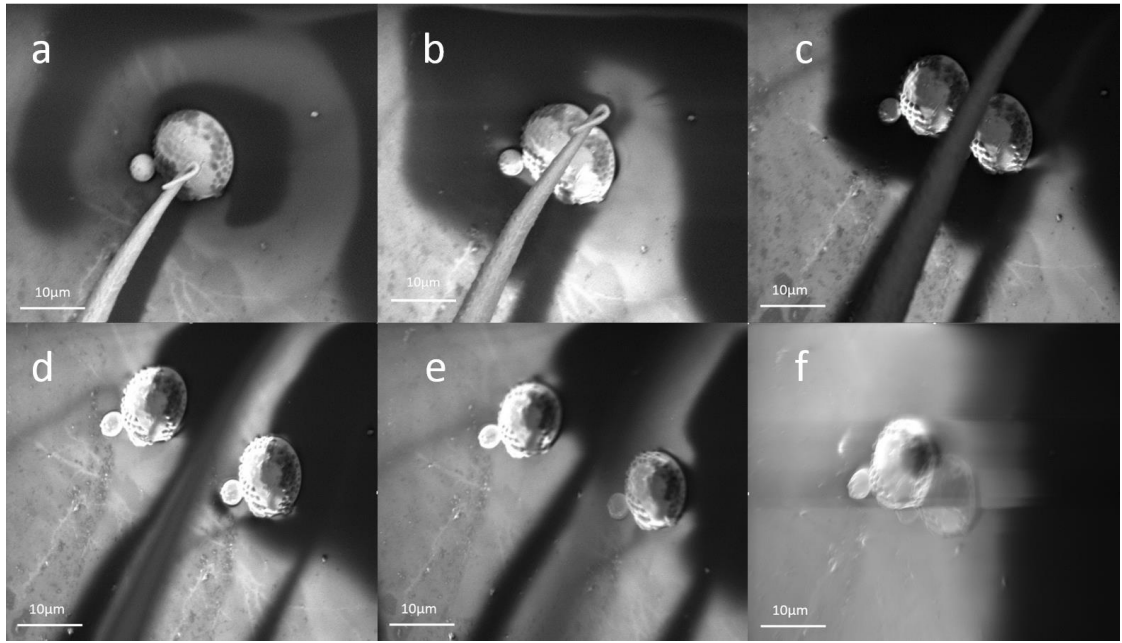

Fig. 3. SEM image of sample with probe height change.
From (a) to (f): the wolfram probe Elevated and away from the particle.

**Discussion**

It is known that insulator sample will lead to anomalous contrast when observed in SEM due to the charging effect, so this phenomenon is also related to the charging effect. Since the substrate was glass which is an insulator material, the surface of the substrate get charged after irradiated by the incident electron beam, and the hemispherical sample also trapped lots of charges. Many works on the influence of charging effect have been reported before[8-10], however, this phenomenon have not been report before, the notable difference between previous reports is that this phenomenon was carried out with a controllable wolfram probe. In the condition of charged specimen, the internal electric field broke by the added probe, thus the double image appeared.

From Figure 1 to Figure 3, anomalous contras occurred when the insulator sample got charged by incident electron beam deflection, while, with the probe added, the image of sample duplicate to a double image, this change caused by the probe is obviously. When the surface of sample gets charged, the internal electric field generated, compare to the size of substrate and sample, the electric field can be treated as parallel electric field, however, since the shape of protruding particles, large quantities of charges were centralized to the surface of particles. When the sample was observed with common condition in SEM, normal charging effect occurred as shown in Figure 1, while, after the metal probe added in, internal electric field changed by the probe and lead to the double imaging.

It is known that induced charges generated when metal conductor placed in electric field, moreover the induced electric field would affect the incident electron beam, distorted incident electron beam lead to imaging distortion, then double image generated on the screen. As it is shown in Figure 4, by electromagnetism knowledge, wolfram probe in SEM can be treat as a conductor in electric filed when it close to the sample, as the probe is electrical grounding, the induced charge on the surface of probe is positive charge, and electric potential is zero, thus the charges can be calculate by the electromagnetism knowledge. The potential of the center probe (U) is equal to the potential induced by charges on the sample (Q) and probe surface (q):

$$U=\int_{q'} \frac{dq'}{4\pi\varepsilon_0 r} + \int_{Q'} \frac{dQ'}{4\pi\varepsilon_0 R} = 0 \qquad (1)$$

Then:

$$q = \frac{Qr}{R} \qquad (2)$$

Where, $\varepsilon_0$ is the electrical permittivity of the vacuum, r is the probe radius, R is the distance between sample and probe center, q' and Q' are the unit charge per area of the surface of probe and sample.

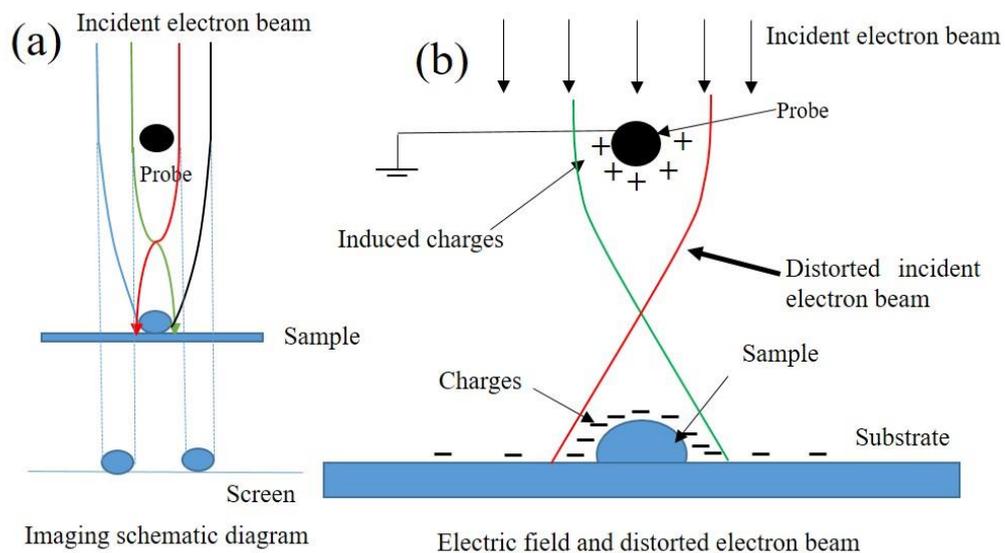

Fig 4. Double imaging schematic diagram

From the equal (2), the charges on the surface of probe is related to the charges on the sample and distance between them, that is the potential on the surface of probe is positively related to charges of sample and negatively related to distance. As a result, the double imaging phenomenon is understood. The influence of probe to the electric field is shown in Figure 4 (b), when the probe is on the top of sample, the incident electron beam distorted by the induced electric field, that is to say the electron beam shift to another place instead of the place it ought to be when it landed to the sample. According to previous articles[10, 11] estimation of the deflection can be tens of microns while the diameter of particles is range from 10-20 microns. The double imaging schematic diagram is shown in Figure 4 (a), when the probe placed in a proper location, the incident electrons deflect to another place as shown with solid line, that is to say the electrons scanned the sample particle twice, as a result the double image generated on the screen in different place as shown in dash line. When the probe did not placed in a symmetrical location, the image is distorted as shown in Figure 2, the image of the sample changed from distort to double. When the probe is connected to sample, shown in Figure 3 (a), the image is ordinary as the probe is electric grounding and charges flow to ground, the charging effect is disappeared. With the height between probe and sample increase, the incident electron was affected by induced charges and the deflected distance located in sample changed by the height, as shown in Figure 3 (b)-(f), notably, in the Figure 3 (f), the double image become closely and unclearly since the height is not only affect the induced charges but also the deflected distance, the higher the less charges, less charges will lead to less deflection, while more height will have more move time to deflection, it is an interaction to the deflection.

**Conclusion**

Double imaging phenomenon was observed in SEM with an added wolfram probe, and it is the first time the phenomenon have been reported and explained. Different from some anomalous contrast reported in SEM, this phenomenon is not only caused by charging effect but also influenced by the induced charges caused by the added wolfram probe. The added wolfram probe changed the surface electric field of the sample under the charging effect and induced charges generated on the surface of probe under the condition of electrical grounding, as a result, incident electron deflected and resulting in double imaging.

In this work, the influence of added probe to imaging in SEM have been studied, when insulator materials observed in SEM charging effect occurred and lead to a surface electric field, with the addition of probe charges distribution changed and influent the incident electron ,thus double image appeared. The internal electric field in SEM is complicated, double imaging is a deduced explanation, more work should be taken to get a more accuracy and understanding of this phenomenon

Acknowledgments

Thanks to Professor Ji Yuan and Wang Li (Beijing University of Technology) for their help of SEM and CL experiments.